\documentclass[aps,pre,preprint,groupedaddress]{revtex4-1}

\usepackage{graphicx}
\usepackage{color}
\usepackage{amssymb,amsmath}
\usepackage{epstopdf}
\usepackage[colorlinks,urlcolor=blue,citecolor=blue,linkcolor=blue]{hyperref}
\usepackage{centernot}
\usepackage[normalem]{ulem}
\usepackage[fleqn]{mathtools}
\usepackage[fleqn]{amsmath}

\begin{document}


\title{Cross-field transport in Goldreich-Sridhar MHD turbulence}


\author{F. Fraschetti}
\email[]{ffrasche@lpl.arizona.edu}
\affiliation{Departments of Planetary Sciences and Astronomy, University of Arizona, Tucson, AZ, 85721, USA}


\date{\today}

\begin{abstract}
I derive analytically the temporal dependence of the perpendicular transport coefficient of a charged particle in the three-dimensional anisotropic turbulence conjectured by Goldreich-Sridhar by implementing multi-spacecraft constraints on the turbulence power spectrum. The particle motion away from the turbulent local field line is assessed as gradient/curvature drift of the guiding-center and compared with the magnetic field line random walk. At inertial scales much smaller than the turbulence outer scale, particles decorrelate from field lines in a free-streaming motion, with no diffusion. In the solar wind at $1$ AU, for energy sufficiently small ($< 1$ keV protons), the perpendicular average displacement due to field line tangling generally dominates over two decades of turbulent scales. However, for higher energies ($\simeq 25$ MeV protons) within the range of multi-spacecraft measurements, the longitudinal spread originating from transport due to gradient/curvature drift reaches up to $\simeq 10^\circ- 20^\circ$. This result highlights the role of the perpendicular transport in the interpretation of interplanetary and interstellar data.
\end{abstract}


\maketitle

\section{Introduction}

The diffusion of charged particles in a turbulent magnetic field plays a pivotal role in the understanding of the origin of cosmic rays, more than a century since their discovery, over a broad range of particle energy, from the interplanetary solar energetic particles \citep[hereafter SEP]{g13} to the PeV cosmic rays likely produced at individual supernova remnant shocks \citep{r08}. In a volume of space (for instance interplanetary region close to the Sun, or interstellar medium stirred by supernova explosion) threaded by a strong statistically uniform average magnetic field $B_0$ with a small fluctuation $\delta B$, the particle diffusion parallel to the average field has been originally modeled in \cite{j66,e74}. The discrepancy later found \citep{p82} between the mean-free path of some SEP events and the prediction of the quasi-linear theory (hereafter QLT) spurred intense theoretical work. \citet{d93} has shown that including the measured, despite noisy, steepening of the power spectrum at high wavenumber for selected events approaches the predicted mean free path to the measurements. \citet{b94} concluded that the geometry of the magnetic fluctuations in the solar wind must be different from the simple picture of QLT. 

Perpendicular diffusion is usually regarded as negligible compared to the parallel diffusion. Nevertheless, the interpretation of a number of heliospheric measurements and numerical simulations of energetic particles suggests a significant contribution arising from perpendicular transport. Time-intensity profiles of energetic protons at spacecraft separated by more than $180^\circ$ in longitude at $1$ AU provide evidence of a significant motion across the Parker spiral magnetic field lines \citep{d14}. Whether the motion across the spiral is due to a pure field line meandering or, in addition to that, to the departure of protons from the actual local field lines, called here cross-field diffusion, is among the purposes of this investigation. From the $X$-ray variability in the solar flares observed by RHESSI \citep{khb11} it has been argued that perpendicular chaotic motion of a few tens of $keV$ electrons in flaring loops was  observed in a non-diffusive phase. Test-particle simulations \citep{fg12} show that even in a weak three-dimensional isotropic turbulence charged particles decorrelate from magnetic field lines on a time scale comparable to the gyroperiod. These results pinpoint to a propagation regime wherein early-time perpendicular transport cannot be neglected.

It is well known that the transport of charged particles in turbulence depends on the anisotropy of the power spectrum \citep[chap. 13]{s02}. Anisotropic turbulence in the solar wind was first found by \citet{bd71} as elongation of turbulence fluctuations along the magnetic field, and confirmed by several studies later on \citep[e.g.]{oh07}. High-heliolatitude {\it Ulysses} measurements \citep{hfo08} provided a new insight: the power spectrum of the local magnetic field in the solar wind is consistent with the anisotropic incompressible MHD-scale turbulence conjectured by \citet{gs95} (hereafter GS95). Such an interpretation was confirmed over a frequency interval comparable with the presumably ``entire''  inertial range of the fast solar wind turbulence \citep{whcs10}. 
A scrutiny of the power spectrum dependence on the angle of the flow to the magnetic field and on the wave-vector anisotropy \citep{fwh11} confirmed an approximate consistency with GS95. \citet{fwh11} argue that the disagreement in the outer length-scale predicted by GS95, inferred to be $20$ times larger than measured, might be explained with unbalance between Alfv\'en modes propagating in opposite directions, in contrast with the balance assumed in GS95. However, the GS95 model does not account for the difference in power-law exponents of the power spectrum of velocity and magnetic  fluctuations in the inertial range measured, e.g., by Wind throughout the solar cycle 23 \citep{p07}. Numerical simulations of MHD turbulence have pointed out that the anisotropy increases at small scales \citep{smm83} and the turbulent eddies become more elongated along the direction of the local magnetic field \citep{cv00,mb00,mg01}.

In this paper I investigate the time-dependent perpendicular transport within the MHD-scale turbulence GS95. \citet{c00} pioneered the study of pitch-angle scattering in the GS95 turbulence within the QLT limit \citep{j66}, finding a scattering frequency more than ten orders of magnitude smaller than the isotropic turbulence over $4$ decades of particle kinetic energy. Extension to a regime immune to QLT divergence at small pitch-angle \citep{v75} confirmed such a scattering inefficiency \citep{yl08}. Since in the GS95 anisotropic turbulence the wave-vectors of the MHD-scale fluctuations are predominantly perpendicular to the local average field, one could expect a small ratio of the perpendicular to the parallel average displacement as compared to the 3D isotropic case. Note that test-particle simulations by \citet{ldkm13} found that the ratio of perpendicular to parallel diffusion coefficient in a synthesised scale-dependent anisotropic turbulence devised to reproduce GS95 is a few percent, comparable to the isotropic turbulence.

Diffusion perpendicular to the mean magnetic field has been long known to be dominated by the meandering of the magnetic field lines \citep{j66}. The rate of separation of turbulent field lines along the direction of the mean field was found in \citet{j73}, by solving a Fokker-Planck equation, to be very fast in the 3D isotropic power-spectrum turbulence, resulting in a significant perpendicular transport solely due to field-line meandering. \citet{rr78}, arguing that field lines diverge exponentially along the mean field direction, found that even if the particle does not undergo significant pitch-angle scattering (``collisionless'' case), the jump from a field line to another warrants a strong perpendicular diffusion due to the exponential divergence. \citet{cc98} showed that an exponential divergence of field-lines reduces the electron thermal conductivity in the galaxy clusters medium. \citet{cm04} used a variety of methods, including Fokker-Planck equation for a static GS95 MHD turbulence and a numerical Monte-Carlo model, and found a field-line separation growing as a power-law. Such theoretical analyses omitted gradient/curvature drift motion of the guiding-center; this assumption is justified for the thermal electrons in the galaxy clusters medium. The present paper investigates the limits of neglecting the gradient/curvature drift in the solar wind turbulent plasma.

In this paper I use the decomposition of the instantaneous perpendicular average square displacement of a charged particle in a magnetic turbulence in two distinct contributions: meandering of magnetic field lines and gradient/curvature drift of the particle guiding-center from the local field line. Those have been analytically calculated for the slab- and the 3D isotropic turbulence in \citet[hereafter FJ11]{fj11} and the latter numerically confirmed for the 3D isotropic turbulence \citep{fg12}. The gradient/curvature drift is usually neglected in theoretical analysis or interpretation of numerical simulations as the cross-field motion is assumed to be dominated by jumps between field lines once the particle has travelled a certain distance away from the original field line. In this paper I calculate gradient/curvature drift and field line contributions to the perpendicular transport for the GS95 anisotropic turbulence and investigate the departure of particles from the field lines as a function of time. Benchmark parameters are tailored to solar wind energetic particles events.

This paper is organized as follows: in Sect. \ref {anisotropy} the power spectra of Alfv\'en polarisation modes relevant to perpendicular transport in GS95 turbulence are described; in Sect. \ref{coeff_diff} the calculation of the instantaneous transport coefficients, due to gradient/curvature drift and to magnetic field line random walk, is outlined (details are in the appendices) and compared with the previous finding of an exponential divergence of field lines; in Sect. \ref{Discussion} the two contributions to the perpendicular transport are compared and the role of gradient/curvature drift average displacement in the SEP longitudinal spread is estimated; Sect. \ref{Conclusion} contains the conclusions.

\section{\label{anisotropy}Anisotropic power spectrum}

Critical balance condition along with the assumption that the energy cascade rate is scale-independent (GS95) yields the scaling $k_\parallel \sim k_\perp^{2/3}/L^{1/3}$, where $k_\parallel ,  k_\perp$ are the components of the wavenumber ${\bf k }$ parallel and perpendicular to the local average magnetic field, respectively, and $L$ is some outer scale which might coincide with the injection scale of the turbulence. Such a scaling holds for a large-scale field Alfv\'enic Mach number $M_A \simeq 1$, as originally assumed in GS95; a more general scaling reads $k_\parallel \sim  k_\perp^{2/3}/L^{1/3} M_A^{4/3}$ \citep{lv99}. The previous scaling implies that the anisotropy increases at the smaller scales. The power at wave-vectors not satisfying the critical balance is exponentially suppressed. However, as well known, the functional form of the power spectrum is not predicted by the GS95 conjecture and various possibilities have been explored \citep[e.g.]{c00}. 

In the GS95 anisotropy the scales of the turbulent eddies parallel and perpendicular are measured only with respect to the direction of the local magnetic field \citep{lv99} that varies according to the location and the scale, whereas in slab or 3D isotropic weak turbulence the direction of the global average statistically uniform field is independent on the scale. Numerical simulations confirm the ratio of parallel and perpendicular scales of the eddies, as well as the relation between $k_\parallel$ and $k_\perp$, implied by the GS95 power spectrum \citep{cv00,mg01}. 

A spatially homogeneous, fluctuating, time-independent global magnetic field ${\bf B}_0$ is considered. The total field is decomposed as ${\bf B(x) = B}_0 + \delta {\bf B(x)}$, with an average component ${\bf B}_0 = B_0 {\bf e}_z$ and a fluctuation $\langle \delta {\bf B(x)} \rangle = 0$. The unit vector ${\bf e}_z$ gives the direction of the global average field component. 
Such a time-independent decomposition of ${\bf B}$ applies to the solar wind, where the propagation of the magnetic fluctuation is much smaller than the velocity of the bulk ionized fluid.

\begin{figure}
\includegraphics[width=13cm]{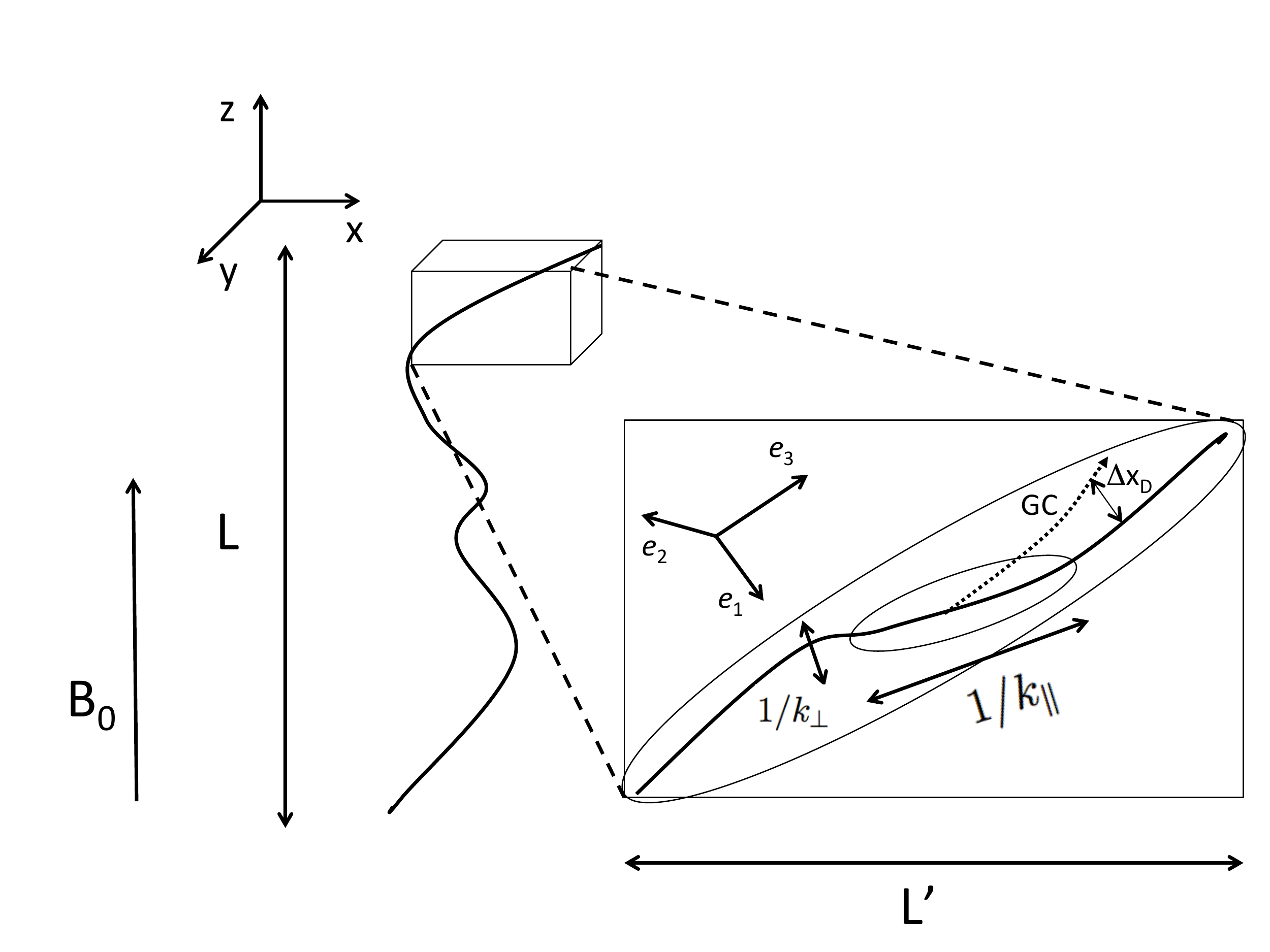}
\caption{Direction of the local average magnetic field, given by the unit-vector $e_3$ at scale $L'$. ``GC'' labels here the trajectory (dotted) of the guiding center drifting away from the local field line due to gradient/curvature drift by the average displacement $\Delta x_D = \sqrt{\langle \Delta x_D^2 \rangle}$ within an elapsed time $\Delta t$ (see Sect. \ref{sect_drift}). } 
\label{Cartoon}
\end{figure}

In this paper, we assume that the inertial range extends from the outer scale $L$ down to some scale wherein injected energy ultimately must dissipate. We introduce a scale $L' \ll L$, much larger than the gyroradius $r_g$ of the particles considered here (Fig. \ref{Cartoon}). The fluctuations  $\delta {\bf B(x)}$ are large compared to the total magnetic field at scale $\sim L$; however, at scales $<  L'$ the magnetic fluctuations are small compared to the local average field, and the first-order orbit theory applies to eddies of scale $< L'$. By using such assumptions, \citet{c00} applied QLT to determine the scattering frequency in the turbulence GS95. The resulting parallel transport depends only on the large scale $L$ and not on the intermediate scale $L'$. 

The most general form of the power spectrum tensor for MHD fluctuations was derived in \citet{orm97}, in their Eq. (20), and in the assumption of axisymmetric turbulence (neglecting cross-helicity) it reduces to 
\begin{equation}
P_{ij} ({\bf k}) =  \left(\delta_{ij} - \frac{k_i k_j}{k^2} \right) E({\bf k}) + \left[ (e_i k_j + e_j k_i)({\bf e}\cdot{\bf k}) - e_i e_j k^2 - \frac{k_i k_j}{k^2} ({\bf e}\cdot{\bf k})^2 \right] F({\bf k})
\label{Oughton_PS}
\end{equation}
where ${\it e}_i$ is the unit vector in cartesian coordinates and ${\bf k}$ the wave-number ($i,j = 1,2,3$), where $k_3 \equiv k_\parallel$ (${\bf k}_\perp = (k_1, k_2)$) is the component of the wavenumber vector parallel (perpendicular) to the local average field (the unit vector $e_3$ is not parallel to the z-axis, see Fig. \ref{Cartoon}). The scalar functions $E({\bf k}), F({\bf k})$ are interpreted in terms of different linear MHD polarisation modes for an oblique orientation between ${\bf k}$ and the local average magnetic field. In particular, $E({\bf k})$ is related to the power in the shear-Alfv\'en modes $\Sigma({\bf k})$, i.e., $\Sigma({\bf k}) = E({\bf k}) $, and $F({\bf k})$ to the difference between shear-  and pseudo-Alfv\'en modes power $\Pi({\bf k})$, i.e., $F({\bf k})  = (\Sigma({\bf k}) - \Pi({\bf k}))/k_\perp^2 $. 

The power spectrum given by Eq. \ref{Oughton_PS} includes the shear- and pseudo-Alfv\'en mode only, and assumes vanishing cross-helicity, i.e., the modes $C({\bf k})$ and $H({\bf k})$ in Eq.(20) in \citet{orm97}. Fast solar wind measurements \citep{wfho12} are compatible with vanishing mode $C$; here the mode $H$ is neglected, as expected to be comparably small. 

A series of incompressible MHD numerical simulations \citep{clv02} implemented the axisymmetric power spectrum tensor given by Eq.\ref{Oughton_PS} to empirically determine the functional form of $E $ and $F$ and best-fit the shear-modes power $\Sigma({\bf k}) $ by using 
\begin{equation}
{\Sigma(k_\parallel, k_\perp) = \frac{{\cal N}B_0^2}{ L^{1/3}} k_\perp^{-10/3} {\rm exp}\left( -\frac{L^{1/3} k_\parallel}{k_\perp^{2/3}}\right)}
\label{shear}  
\end{equation}
where ${\cal N}$ is a normalisation constant to be determined, $2\pi/  L' < k_\perp < k_ \perp ^M $, where ${k_\perp^M}$ is the power spectrum cut-off where dissipative effects start to be important and MHD approximation breaks down, and $M_A \simeq 1$. According to the GS95 conjecture, the pseudo-Alfv\'en modes are carried passively by the shear-Alfv\'en modes (this property is an exact result for weak MHD turbulence \citep{g00}) with no contribution to the turbulence cascade to small scales which is seeded by collisions of shear modes only. Thus, it is reasonable to assume that $\Pi_{ij}({\bf k}) $ has the same functional form as $\Sigma_{ij}({\bf k}) $, allowing for a relative amplitude $\varepsilon$ ($0 < \varepsilon < 1$) and a different outer scale $\ell$ ($\ell \neq L$) both so far inaccessible to either solar wind observations or models. Likewise, the power spectrum of the pseudo-modes is cast in the form
\begin{equation}
{ \Pi (k_\parallel, k_\perp)  =  \frac{\varepsilon {\cal N} B_0^2}{ \ell^{1/3}}  k_\perp^{-10/3} {\rm exp}\left( -\frac{\ell^{1/3} k_\parallel}{k_\perp^{2/3}}\right)} .
\label{pseudo}  
\end{equation}
The constant ${\cal N}$ is fixed by normalisation of both polarisation modes that can be extended to scales between $L'$ and $L$, as the contribution from scales between $L'$ and $L$ is exponentially suppressed. Within the range $k_{\parallel , \perp} ^0 \ll k_{\parallel , \perp} \ll k_{\parallel , \perp} ^M $, if the largest scales $1/k_\parallel ^0, 1/k_\perp^0$ are related to the outer scale by $k_\parallel ^0 \simeq (k_\perp^0)^{2/3} / L^{1/3}$, the normalisation reads: $\delta B ^2 = \int d^3 k ~ {\rm Trace}(P_{ij}) = \int d^3 k (\Sigma ({\bf k}) + \Pi ({\bf k}))$ which gives ${\cal N} \simeq {1\over 3\pi} (\delta B/B_0)^2 (1 + \varepsilon (L/\ell))^{-1}$, where $k_\parallel ^0 L \simeq 1$ was used. We note that ${\cal N}$ does not depend on the scale $L'$ neither on $k_{\parallel} ^M$, that is irrelevant as the cascade along the local average field is suppressed.

Recent analysis of fast solar wind data from {\it Ulysses} \citep{wfho12} provides us with an estimate of $\varepsilon$ by confirming an ordering in diagonal components of the power tensor that has long been found \citep{bd71}: if $e_3$ is the unit vector in the direction of the local average magnetic field, the measurements give a power along $e_3$ smaller, but not negligible ($\varepsilon \centernot\ll 1$), than in the plane orthogonal to $e_3$. As shown in Sect. \ref{coeff_diff}, such a power ordering results in a small but not negligible effect on the perpendicular transport of the pseudo modes with respect to the shear modes. The forms given by Eq.s \ref{shear}, \ref{pseudo} are used in Sect. \ref{coeff_diff} to calculate the time-dependent perpendicular transport. 

\citet{c00} used a power spectrum with steep cutoff beyond the critical balance scaling $k_\parallel \sim k_\perp^{2/3}$ to determine the scattering frequency within the QLT approximation. In this paper I calculate the time-dependent perpendicular transport coefficient by expanding in Taylor series the exponentials ${\rm exp}( -L^{1/3} k_\parallel/k_\perp^{2/3})$, $  {\rm exp}( -\ell^{1/3} k_\parallel/k_\perp^{2/3})$ in the power spectra in Eqs. \ref{shear}, \ref{pseudo}; such an expansion is valid as the power spectrum is a convergent quantity. Only the terms growing fastest in time are retained. We note that no assumption is necessary on the explicit spatial dependence of $\delta {\bf B} ({\bf x})$, that has three non-vanishing space components, as only the Fourier transform $\delta {\bf B} ({\bf k})$ is used.

\section{\label{coeff_diff}Early-time perpendicular transport coefficient}

The magnetostatic approximation has been widely used for decades \citep{j66} to investigate the transport of charged particles in magnetic turbulence, as the speed of the particles considered is typically much greater than the average-field Alfv\'en speed. Solar wind flows outward from the Sun at a speed of several hundreds of km$/$s (300-400 at low heliolatitudes and 700 at high heliolatitudes), much faster than the local Alfv\'en speed (tens of km$/$s) and the spacecraft speed (a few km$/$s). Thus, the magnetic fluctuations can be assumed to be frozen with the plasma. In this paper I investigate the perpendicular transport at time-scales shorter than the correlation time of the perpendicular fluctuations of the local magnetic field, as seen by the particle. The transport perpendicular to the average magnetic field is dominated by guiding-center motion which includes the meandering of the magnetic field lines (MFL) and the gradient/curvature drift from the first-order orbit theory \citep[FJ11]{ro70}, in the approximation that the particle gyroradius is much smaller than the scale of magnetic field variations. The existence of a diffusion regime for the perpendicular transport is not assumed.

\subsection{\label{sect_drift}Gradient/Curvature Drift Transport}

In this section, the derivation of the gradient/curvature drift transport coefficient in FJ11 is outlined and the result in the case of the GS95 turbulence is presented (details in Appendix A). The first-order orbit theory \citep{ro70} is used, assuming that $r_g$ is much smaller than the length-scale of any magnetic field variation: $r_g \ll \displaystyle\min_{i,j=1,3} |B_i/\partial_j B_i|$  or, in other terms, the magnetic field varies slightly at the gyroscale so that the particle orbit can be approximated within a gyroperiod by the helicoidal trajectory with local curvature radius given by $r_g$; for wave-numbers $k_\perp$ within the inertial range, the smallness of $r_g$ reads $k_\perp r_g \ll 1$; in other terms we assume that the perturbation in the turbulent cascade at the gyroscale $r_g$ does not affect significantly the particle trajectory.

\begin{figure}
\includegraphics[width=8.cm]{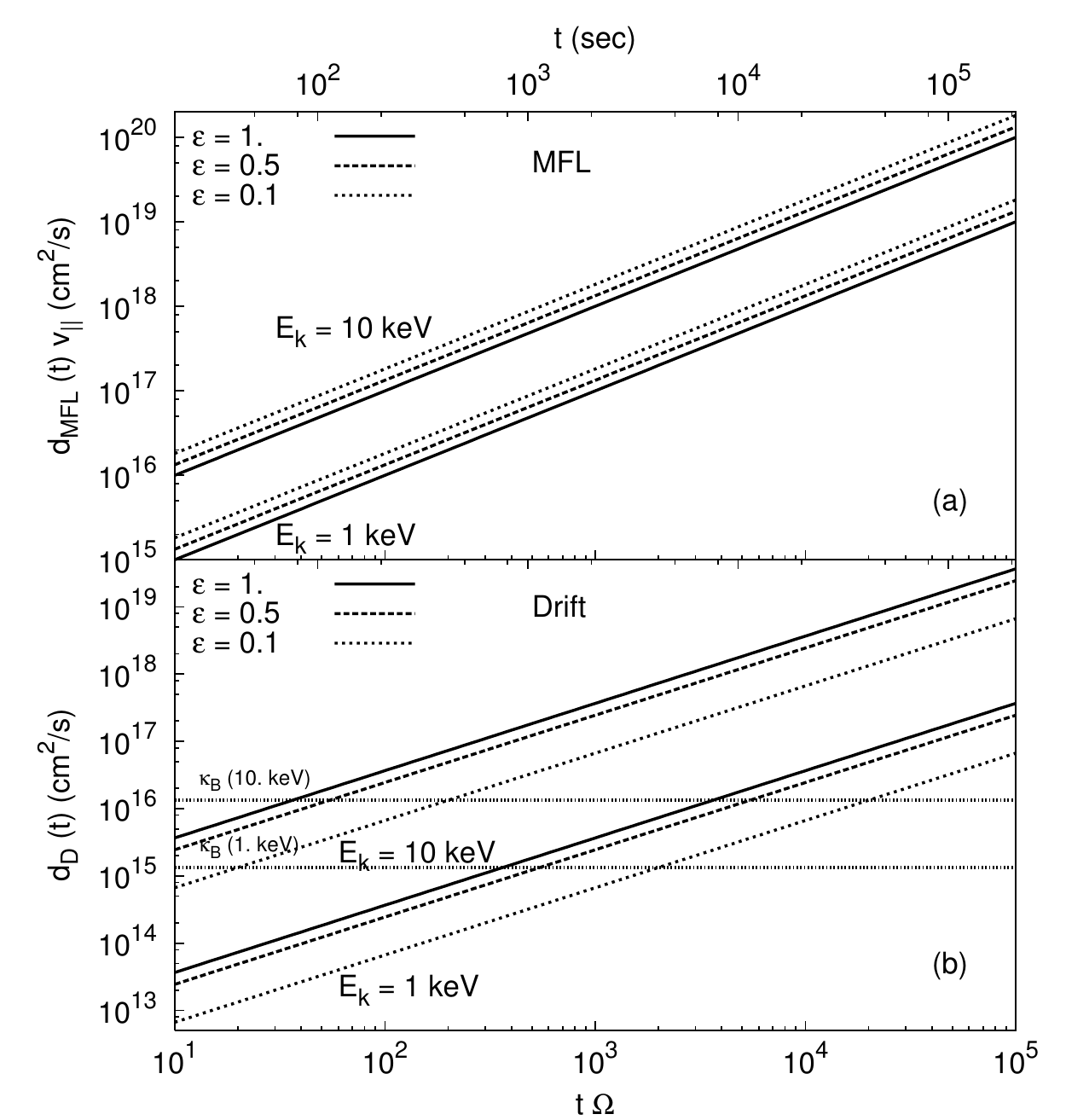}
\includegraphics[width=8.cm]{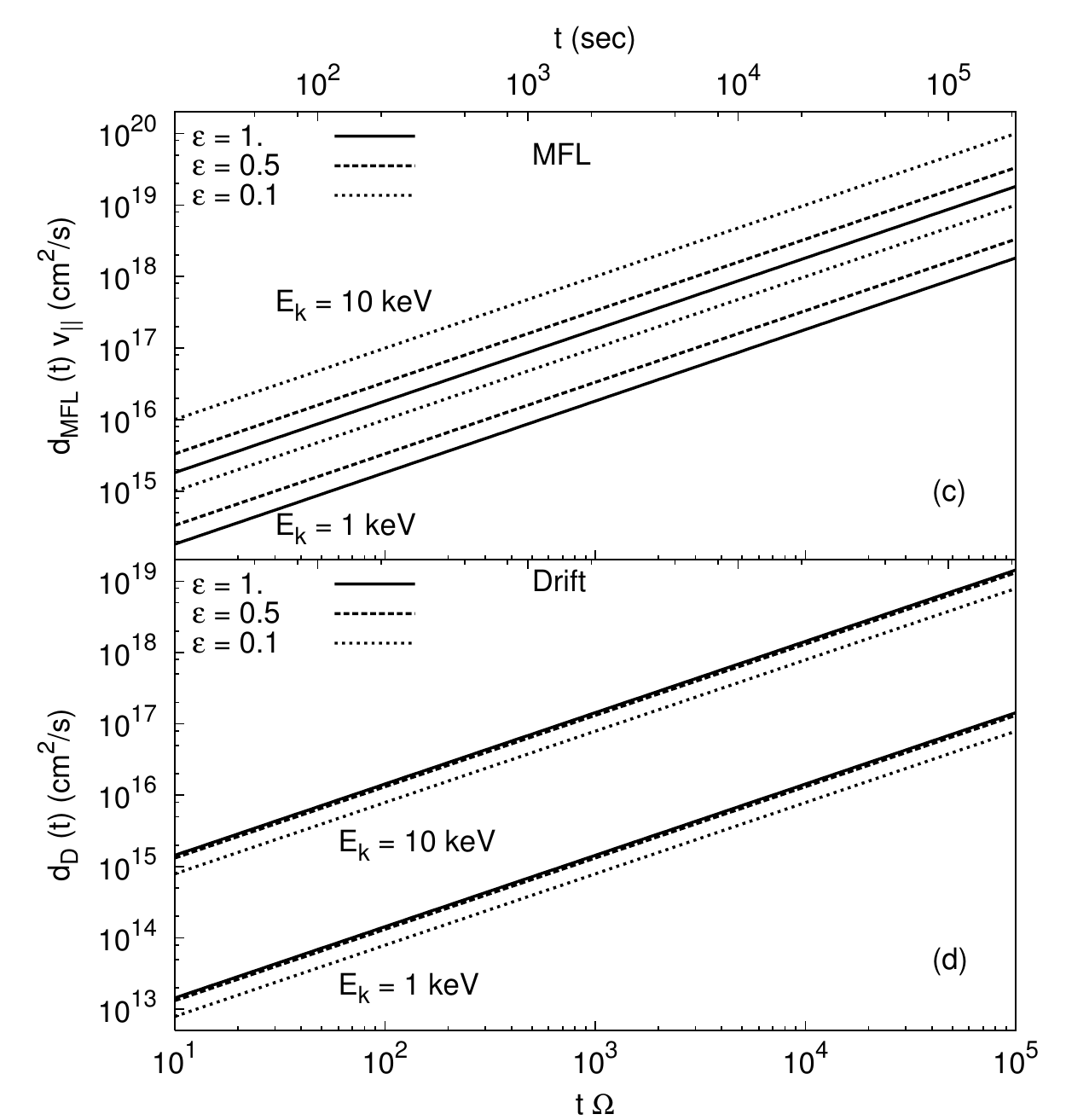}
\caption{Magnetic field line (a,c) and gradient/curvature drift (b,d) instantaneous perpendicular average square displacement as a function of time for protons with kinetic energy $E_k = 1$ keV and $E_k = 10$ keV in the solar wind at $1$ AU with $L/\ell = 1$ (a,b) or $L/\ell = 10$ (c,d); the solid (dashed, dotted) line corresponds to $\varepsilon = 1$ ($\varepsilon = 0.5$, $\varepsilon = 0.1$); also $B_0 = 5$ nT, $(\delta B /B_0)^2 = 0.1$, $L = 0.01$ AU ($r_g /L = 6.1 \times 10^{-4} , 1.9 \times 10^{-3}$ respectively) and $k_\perp^M L = 10^3$. The horizontal lines in (b) correspond to the Bohm diffusion at the particle energy considered.}
\label{fig_time}
\end{figure}

At scales $< L'$, we calculate the gradient/curvature drift from the local average field at those scales. We make use of the gyroperiod averaged guiding-center velocity transverse to the local field ${\bf B (x)}$, i.e., ${\bf V}_{\perp}^G (t)$, to the first order in the fluctuation, given for a particle of speed $v$, momentum $p$ and charge $Ze$ by \citep{ro70}
\begin{equation}
{\bf V}_{\perp}^G (t) =  \frac{vpc}{Ze B^3} \left[ \frac{1+\mu^2}{2} {\bf B}\times \nabla B + \mu^2 B (\nabla \times {\bf B})_\perp   \right] ,
\label{Vperp}
\end{equation}
where $\mu$ is the cosine of the pitch-angle with respect to the local average field. The average square transverse displacement of the particle from the direction of local field due to drift, $d_D (t)$, at time $t$ is written as
\begin{equation}
d_{D} (t) = \int _0 ^t  d\xi \langle {\bf V}_{\perp, i}^G (t') {\bf V}_{\perp, i}^G (t' + \xi) \rangle  .
\label{dXX1}
\end{equation} 
The drift transport coefficient would be defined here as the limit $\kappa_{D} = \displaystyle\lim_{t \rightarrow \infty} d_{D} (t) $. We assume that $P_{ij}({\bf k})$ is uncorrelated at different wave-number vectors, i.e. $\langle \delta B_r({\bf k}) \delta B_q ^*({\bf k'})  \rangle = \delta ({\bf k} - {\bf k'}) P_{rq}({\bf k})$ for the Fourier components $\delta B_r({\bf k})$; we note that the validity of the critical balance condition scale-by-scale used here has been recently questioned in a phenomenological analysis \cite{m14} that reports inconsistencies with ion kinetic scales for parameters in the solar wind regime. The generic non-vanishing term of Eq.\ref{dXX1} (regardless of the power spectrum assumed) is given by (see FJ11):
\begin{equation}
\left(\frac{vpc}{Ze B_0 ^2} \right)^2 F(\mu^2) \int_{-\infty} ^{\infty} d^3 k  P_{rq}({\bf k}) k_l k_p \frac{{\rm sin} [k_\parallel v_\parallel t]}{k_\parallel v_\parallel} \, ,
\label{Fmu2}
\end{equation}
with indexes $(r,q,l,p) = (3,3,2,2)$, $(3,2,2,3)$, $(2,3,3,2)$, $(2,2,3,3)$ for $d_{D_{XX}}$ and $(r,q,l,p) = (3,3,1,1)$, $(3,1,1,3)$, $(1,3,3,1)$, $(1,1,3,3)$ for $d_{D_{YY}}$. In the factor $(vpc/Ze B_0 ^2)^2$ in Eq. \ref{Fmu2} we assumed that the magnitude of the local average field can be approximated by $B_0$ at those scales. Here $F(\mu^2)$ represents various factors depending on the particle pitch-angle $\mu$ resulting from the expansion of Eq.\ref{dXX1} (FJ11). 
Such a calculation is detailed in Appendix \ref{app_drift}. From a comparison of the terms in Eqs. \ref{series_tot_1} , \ref{series_tot_2}, \ref{series_tot_3} and \ref{series_tot_4}, only the term fastest growing in time (Eq.\ref{series_tot_1}) is retained: the drift transport coefficient is dominated by the power of the pseudo-Alfv\'en modes along the local average field $P_{33} ({\bf k})$ (term $3322$ in Appendix A).  Thus, by using Eq.\ref{Fmu2}, the instantaneous drift transport coefficient can be cast as
\begin{align}
d_{D} (t) \simeq \frac{1}{60} \left(\frac{\delta B}{B_0} \right)^2 \frac{\varepsilon}{1+\varepsilon L/\ell}  \left(\frac{L}{\ell} \right)^{1/3}   \left(\frac{ r_g}{L} \right)^2 r_g v {\Omega t}   
\left[ (k_\perp L)^{8/3}  {}_{2}F_{1} \left.\left(1,{4\over 3}; {7 \over 3};- (k_\perp L)^{2/3} \right) \right]\right|_{k_\perp ^{M}} 
\label{series_tot_drift}
\end{align}
where $\Omega = Z e B_0/mc$ is the particle gyro-frequency, $r_g \simeq pc/ZeB_0$. The quantities $\Omega, r_g$ depend on the magnitude of the global average field $B_0$. The corrective factor in Eq. \ref{series_tot_drift} is only the ratio of the power of the global to the local average fields (${\cal O}(1)$). We emphasise that Eq. \ref{series_tot_drift} is valid only as long as particle transport is confined within the scale $L'$. In Fig. \ref{fig_time}, panels (b,d) depict the time-evolution of $d_D (t)$ as given by Eq. \ref{series_tot_drift} for energetic protons in the solar wind for different values of $\varepsilon$ {\bf ($1, 0.5, 0.1$)} and $L/\ell =1$ in (b) and $L/\ell =10$ in (d). Higher power in the pseudo-Alfv\'en modes relative to the shear- (i.e., higher $\varepsilon$) results in an enhancement of $d_D(t)$ (see Appendix A). In analogy with the 3D isotropic turbulence case (FJ11), $d_D (t)$ in the anisotropic GS95 turbulence grows linearly with $(\delta B/B_0)^2$. As compared with the Bohm diffusion coefficient $\kappa_B = v r_g /3 = 1.3 \times 10^{16}$ cm$^2 /$s for $E_k = 10$ keV, Fig. \ref{fig_time} (b) shows that $d_D(t) \simeq\kappa_B$ within a very short time ($\Omega t \simeq 50$, $L/\ell = 1$ and $\varepsilon = 1$). We emphasise that the average square displacement due to drift (Eq. \ref{series_tot_drift}) does not depend only on large-scale properties of the turbulence ($L$), but also depends non trivially on the smallest scales ($k_\perp ^M$). The result is independent on the scale $L'$ \citep{c00} as the large scales are exponentially suppressed.

We can also conclude that the guiding-center is confined only for short-time to move along the field line. We compare the average square displacement from the local field-line reached by the guiding-center via gradient/curvature drift, i.e., $\langle \Delta x_D^2 \rangle \sim 2 d_D(t) \Delta t $ (see Fig. \ref{Cartoon}) with the squared gyroscale ($r_g^2$). For a proton with $E_k = 1.$ keV in the solar wind at $1$ AU (see Fig. \ref{fig_time} (b)), we find that $\langle \Delta x_D^2 \rangle$ at $\Omega t =100$ ($\varepsilon = 0.1$) is comparable with $r_g^2 = 8.3\times 10^{15}$ cm$^2$. Thus, after a few gyroperiods the actual particle has an average distance from the local field greater than $2 r_g$.

We remind that the gradient/curvature drift transport is calculated here up to scales $L'$, with a given direction ($e_3$) of the local average field at that scale different from the global average field direction  ($z$-axis). Our finding does not rule out that at scales close to the outer scale $L$ the perpendicular transport can turn into diffusive regime. However, accounting for those larger scales requires a spatial dependence of the local turbulence wave-number vectors; this is deferred to a separate work.

\subsection{Magnetic field line}

In this section the method to compute the time-dependent particle transverse transport due to MFL meandering (FJ11) is outlined and the result for the GS95 turbulence is presented. Previous theoretical derivations of MFL meandering in the isotropic \citep{j73} and anisotropic GS95 turbulence \citep{nm01,cm04} focused on the relative divergence between two initially static nearby (down to thermal electron gyroradius scale) field lines along the direction of the average magnetic field. In contrast with this result, magnetic turbulence has been found \citep{lv99,e11} to exhibit a field line separation in agreement with the long argued Richardson scaling $t^{3/2}$ for a pair of particles in purely hydrodynamical turbulent media. Numerical simulations of MHD turbulence in a partially ionised plasma \citep{lvc04} show that the neutral drag has the effect of recovering the exponential field line separation argued in \citet{rr78}. In analogy with FJ11, we focus on the divergence of a single field line from the direction of the local average field. The method of two field lines separation applies to the case of zero average field; on the other hand, being based on the solution of the Fokker-Planck equation, it requires several independent small displacements $\Delta x_{MFL}$ (central limit theorem), thus it requires diffusive regime to be reached, whereas the method FJ11 applies more generally, i.e., prior to the diffusive regime.

We assume that the correlation function of the magnetic fluctuation is homogeneous in space, i.e., the correlation depends only on distances along the particle orbit: $\langle \delta B_x [{\bf x}(x_3^0 + x_3)] \delta B_x [{\bf x}(x_3^0)] \rangle = \langle \delta B_x [{\bf x}(x_3)] \delta B_x [{\bf x}(0)] \rangle$. The mean square displacement of the
MFL orthogonal to the $e_3$ vector can be defined as 
\begin{equation}
d_{MFL} (x_3) = {1\over 2} \frac{d \langle \Delta x^2 (x_3)\rangle }{ d x_3} = \frac{1}{B_0^2}\int_0^{x_3} d {x_3}   \langle \delta B_x [{\bf x}(x_3)] \delta B_x [{\bf x}(0)] \rangle .
\end{equation}
We compute the magnetic turbulence along the unperturbed trajectory of a pseudo-particle travelling with zero pitch-angle and no scattering (FJ11). The mean-square transverse displacement of a MFL corresponding to a distance $x_3 = v_\parallel t$ along the local uniform field  travelled by a such a pseudo-particle can be written as 
\begin{equation}
d_{MFL} (t) =  \frac{1}{B_0^2} \int_{-\infty} ^{\infty} d^3 k  P_{ij}({\bf k}) \frac{{\rm sin} [k_\parallel v_\parallel t]}{k_\parallel} \, .
\label{dMFL2}
\end{equation}
where $P_{ij}({\bf k})$ is the magnetic turbulence power spectrum in the inertial range (Eq. \ref{Oughton_PS}).
The MFL diffusion coefficient is the limit $\kappa_{MFL} = \displaystyle\lim_{t \rightarrow \infty} d_{MFL} (t) v_\parallel$. As in the previous section, in Eq. \ref{dMFL2} we assumed that the magnitude of the local average field can be approximated by $B_0$. The detailed calculation for the GS95 anisotropic power spectrum (Eq.s \ref{Oughton_PS}, \ref{shear}, \ref{pseudo}) can be found in Appendix \ref{app_MFL}.

From Eq.\ref{shear_MFL_0}, the instantaneous transport coefficient due to MFL meandering is dominated by shear-modes and given by 
\begin{equation}
d_{MFL} (t) v_\parallel \simeq \frac{1}{2 }  \left(\frac{\delta B}{B_0} \right)^2 \frac{1}{1+\varepsilon L/\ell} r_g v {\Omega t}  .
\label{shear_MFL}
\end{equation}
Figure \ref{fig_time}, {\bf (a,c)}, depicts the time-evolution of $d_{MFL} (t) v_\parallel$ as given by Eq. \ref{shear_MFL} for protons at various energies in the solar wind and for different value of $\varepsilon$ {\bf ($1, 0.5, 0.1$)} and $L/\ell =1$ ($L/\ell =10$) in the left (right). In analogy to the MFL random walk in isotropic turbulence (FJ11) and the well-known QLT result, $d_{MFL} (t)$ for anisotropic GS95 grows linearly with $(\delta B/B_0)^2$. Alike the drift, the independence on $L'$ arises from the fact that scales between $L'$ and $L$  are exponentially suppressed. 

We find that even at the small time-scales considered here, different magnetic turbulences exhibit different scalings. In the case of GS95 anisotropy the linear growth in time of the MFL average square displacement $d_{MFL} (t) $ is fast as compared to the 3D-isotropic turbulence, wherein a logarithmic growth at large scales and diffusion at small scales are found (see Eq. [49] in FJ11).

With the substitution $ x_3 = v_\parallel t$, Eq. \ref{shear_MFL} gives $d_{MFL} (x_3) \simeq (1/2)  (\delta B/B_0)^2 x_3/(1+\varepsilon L/\ell) $. We note that a linear dependence of $d_{MFL} (x_3) $ on $x_3$ could also be derived empirically from the argument $\Delta x_1 /\Delta x_3 \sim \delta B_1 /B_0 $. By using a direct calculation we determine here the dependence of the proportionality  coefficient on the turbulence parameters, otherwise to be determined empirically, for instance, through numerical simulations.

\section{\label{Discussion}Discussion}

\subsection{\label{disc_drift}Role of the gradient/curvature drift}

\begin{figure}
\includegraphics[width=10cm]{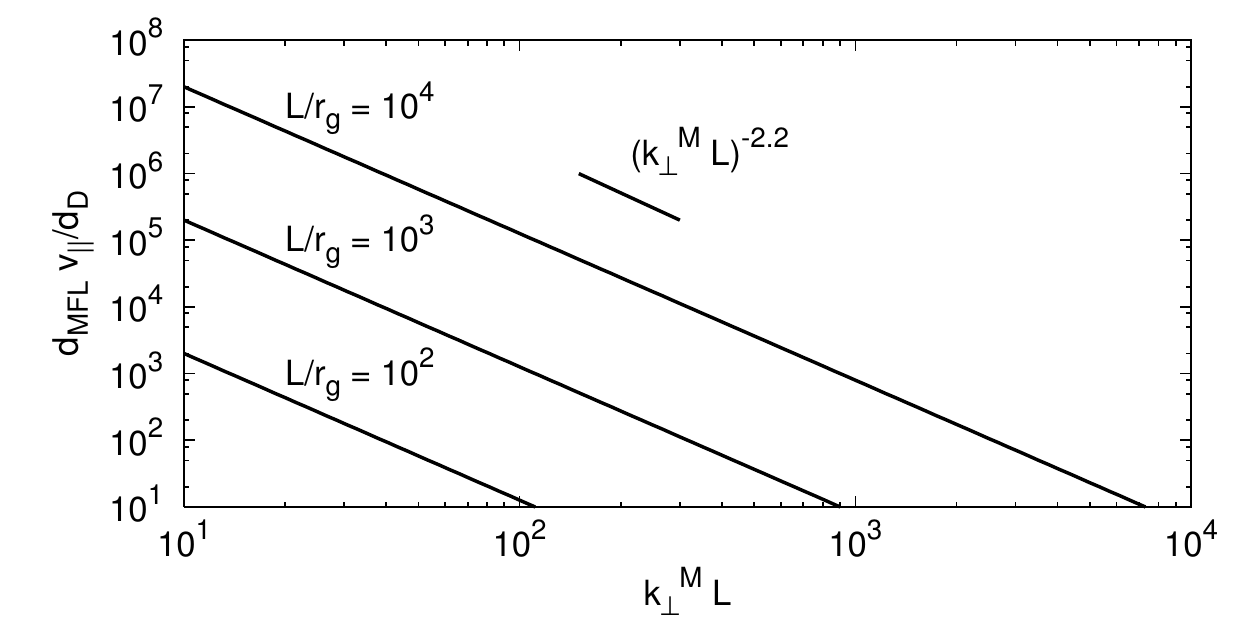}
\caption{Ratio of the MFL perpendicular transport coefficient $d_{MFL} (t) v_\parallel$ to the gradient/curvature drift transport coefficient $d_{D} (t)$, for $L/r_g = 10^2, 10^3, 10^4$, corresponding to proton kinetic energy at $1$AU given by $E_k= 250, 2.5, 0.025$ keV respectively, as a function of $k_\perp^M L$ as determined in Eq. \ref{ratio} (here we used $L/\ell = 1$ and $\varepsilon =1$). } 
\label{fig_ratio}
\end{figure}

We discuss the relative contribution of magnetic field line meandering and gradient/curvature drift to the perpendicular transport in the GS95 anisotropy for times shorter than the correlation time of the perpendicular fluctuations of the local magnetic field, as seen by the particle. In the guiding-center approximation used here, the ratio $d_{MFL}(t) v_\parallel/d_D (t)$ (see Eqs. \ref{series_tot_drift} and \ref{shear_MFL}) is time-independent.  
Such a ratio can be cast as a function of $k_\perp ^M L$ in the form
\begin{equation}
\frac{d_{MFL} v_\parallel}{d_{D }} (k_\perp^{M} L) \simeq 30 \left(\frac{L}{r_g}\right)^2  \left(\frac{\ell}{L}\right)^{1/3}  {1 \over \varepsilon} \left[(k_\perp^{M} L)^{8/3}  {}_{2}F_{1} \left(1,{4\over 3}; {7 \over 3};- (k_\perp^{M} L)^{2/3} \right) \right]^{-1}   .
\label{ratio}
\end{equation}
Figure \ref{fig_ratio} illustrates $d_{MFL} v_\parallel/d_D$ in Eq. \ref{ratio} for different values of $L/r_g$. We emphasise that the ratio $d_{MFL} v_\parallel / d_D$, large in the solar wind, is expected to be upper bounded: $d_D$ grows with the turbulence power along the local average field $P_{33}$, that is small, although not negligible, compared to the other diagonal terms of the power spectrum tensor ($P_{11}, P_{22}$). 

Figure \ref{fig_ratio} shows that $d_D$ becomes relevant at small scales, i.e., at large wave-numbers $k_\perp ^M$. If $d_{MFL}(t) v_\parallel/d_D(t) \sim 10$, the average perpendicular displacements due to tangled field lines and to the gradient/curvature drift become comparable: $\sqrt{d_{MFL}(t) v_\parallel/d_D(t)} \sim 3$. From Fig. \ref{fig_ratio}, we conclude that for a sufficiently small energy ($< 1$ keV) proton at $1$ AU in the solar wind turbulence, using an inertial range extended down to a scale $\sim 10^{-5}$ AU ($k_\perp ^M L \simeq 10^2 -10^3$ for $L = 0.01 $ AU), the average perpendicular displacement due to field line meandering is generally dominant ($d_{MFL}(t) v_\parallel/d_D(t) \simeq 10^5$ or greater). However, for higher energies ($\sim 10$ keV) protons ($L/r_g \simeq  500$), the ratio of the two average square displacements is smaller and within a smaller inertial range ($k_\perp^M L \simeq 10^2$) reaches values down to $\sim 10$. The region $d_{MFL} v_\parallel / d_D < 10$ is inaccessible to our model that applies only to turbulent scale wave-numbers $k_\perp^M$ such that $1 \gg k_\perp^M r_g = (k_\perp^M L) (r_g/L)$. For a given scale $k_\perp ^M$, the ratio $d_{MFL}(t) v_\parallel/d_D(t)$ decreases as the particle energy increases, i.e., for small $L/r_g$, as expected. This result constrains the  customary assumption that charged particles follow turbulent field lines. We note also that the result is independent of the parallel mean-free path. This is in contrast with the model in \citet{sbls10}; however, we remind that the model presented here does not account for parallel transport, that has already been determined to be irrelevant due to very small scattering frequency in GS95 within QLT limits \citep{c00}, and is valid only for times $t < 1/k_\parallel v_\parallel$.

We find that at small scales the perpendicular transport in GS95 turbulence is a free-streaming motion ($d (t) \sim \Delta x ^2/2 \Delta t \sim t$), in contrast with a diffusive regime ($d (t) \simeq const$). Theoretical evidence of perpendicular super-diffusion of the magnetic field lines alone in GS95 anisotropic turbulence ($\Delta x_{MFL} \sim \Delta t ^a$, with $1/2< a < 1$) is discussed also in \cite{ly14}, that focusses on the field line divergence (see references therein), assuming no departure of the particle from the local field line. The fourth-power dependence in $\delta B/B_0$ of the MFL meandering found in \cite{ly14}, in contrast with the second-power found here, is likely to originate from assumptions therein on the turbulence generation and reconnection, i.e., normalisation factor. In addition, we find that the motion of the guiding-center away from the field line might reach, and eventually exceed, the gyroradius scale within a few hundreds $\Omega t$ (as shown in Sect. \ref{sect_drift}) for times smaller than the correlation time of the perpendicular fluctuations of the local magnetic field, as seen by the particle. We note that the departure from field line found here does not violate the reduced dimensionality theorem \citep{jkg93} as no constraint is placed on the ignorable coordinates of $\delta {\bf B}$.

From Eq. \ref{series_tot_drift}, it is readily found (see also Fig.\ref{fig_ratio}) that the drift perpendicular transport coefficient scales with the largest perpendicular wave-number $k_\perp ^M$, i.e., with the smallest perpendicular inertial scales, as $d_D \simeq (k_\perp ^M)^{2.2}$. In GS95 turbulence the anisotropy is known to increase at smaller scales favouring energy cascade in the perpendicular direction. Thus, it is clear that the smaller the scale (namely the larger $k_\perp ^M$), the larger the anisotropy in the power of the turbulence, the more significant the departure from the field lines. This is in agreement with the limit case of the slab-turbulence, i.e., anisotropy wherein the power along the the mean field direction is completely suppressed: the gradient/curvature drift average square displacement $d_D (t)$ is dominated by small scales (FJ11), whereas the larger scales govern the MFL meandering. 

\subsection{\label{SEP_events}Reconstruction of longitudinal spread in SEP events}

In this subsection we provide a crude estimate of the role of the perpendicular transport in the reconstruction of the propagation of SEP from the flaring region out to the multi-spacecraft location at $1$ AU. Strong perpendicular transport has often been invoked to explain the longitudinal spread in multi-spacecraft measurements of SEP events (see \citet{d14} and references therein). According to an alternative plausible mechanism, the longitudinal spread might originate from wide longitudinal extension or motion of CME-driven shocks in the interplanetary medium.

We assume the large-scale ordered field ${\bf B}_0$ to be approximately uniform over a distance much smaller than the putative correlation length scale of the solar wind turbulence ($0.01$ AU). 
Therefore, we can constrain the average distance $\Delta x_D$ associated with gradient/curvature drift that the particles are transported in the direction perpendicular to the local field at various distances from the Sun; we emphasise that such a drift originates from the inhomogeneity of the magnetic fluctuation and not of the spiralling magnetic field itself. The effect of field line meandering is switched off in this paragraph. The field ${\bf B}_0$ is assumed to be piecewise uniform; the resulting cumulative effect might contribute to the longitudinal spread at $1$ AU. 

In the SEP event of January $17^{th}$ $2010$, energetic protons have been measured by the two STEREOs (LET and HET) and SOHO/EPHIN to spread almost over $360^\circ$ at $1$ AU \citep{d12}. LET and HET instruments measure energetic protons in the range from $4$ to $60$ MeV and SOHO/EPHIN from $4$ to $25$ MeV. A possible scenario comprises protons with kinetic energy $E_k$ equal or greater than $4$ MeV that, while travelling along a turbulent field line of the Parker spiral in the ecliptic plane and drifting by some amount away from that line, undergo scattering and back-trace a field line adjacent to the original one. The iteration of such a process over a time interval $\Delta t$ (see table \ref{table}) might spread energetic protons over wide longitudinal angle over a few hours. In this crude estimate, we assume this scattering to be frequent enough that the magnitude $B_0$ can be taken as uniform. The solar wind advection, customarily taken purely radial, is neglected here; we expect it to ultimately only enhance $\Delta x_D$. In table \ref{table} we consider two distances: $1$ AU ($B_0 = 5$ nT) and $10$ solar radii ($B_0 \simeq 0.025$ Gauss). The Sun's spin ($14^\circ/$day) is here neglected. The longitudinal spread $\Delta \alpha$ is calculated in table \ref{table} as $\Delta x_D$ in units of the distance from the Sun, where the average displacement from the field line due to gradient/curvature drift is given by $\Delta x _D \simeq \sqrt{2 d_D (\Delta t) \Delta t} $ and $d_D (\Delta t)$ is defined in Eq. \ref{series_tot_drift}. In table \ref{table} we used $ L =0.01 $ AU$, L/\ell =1$ and, from Eq. \ref{series_tot_drift}, $\left[ (k_\perp L)^{8/3}  {}_{2}F_{1} \left(1,{4\over 3}; {7 \over 3};- (k_\perp L)^{2/3} \right) \right] |_{k_\perp ^{M}L =10} = 149.8$. Table \ref{table} shows that the time elapsed since the particle injection in the turbulent region where $B_0$ can be taken as uniform is a major contributor to the spread $\Delta \alpha$, being the transport ballistic. In particular $E_k = 25$ MeV protons at $\sim 1$ AU after $1$ hour cover a non-negligible longitudinal angle (of order of $10^\circ$). Our estimate is limited by small range of fluctuations interacting with the particles: $25$ MeV protons ($r_g/L = 0.098$ at $1$ AU) is allowed to interact only with one decade of turbulent scales, satisfying the condition $1 \gg k_\perp^M r_g = (k_\perp^M L) (r_g/L)$.  
One can also roughly predict $\Delta x _D$ to be measured by Solar Probe Plus mission in its approach to the Sun at a distance of $10$ solar radii: for $\Delta t = 30$ min, we find $\Delta \alpha \simeq 0.0323^\circ$ for a $E_k = 4$ MeV proton. 

\begin{table}[ht]
\caption{Estimate of the longitudinal spread of energetic protons due to gradient/curvature drift average displacement.}
\centering 
\begin{tabular}{c c c c c c c}
\hline\hline
        $E_k $ (MeV)& $\sigma^2$ & $B_0$ (Gauss) & $\varepsilon$ & $\Delta t$ (hour) & $\Delta x_D$ (AU) & $\Delta \alpha$ ($^\circ$)\\ [0.5ex]     
        \hline     
4    &    0.1 &   $5.\times 10^{-5}$ &   1   &  1    &   $0.0128$  		   & $0.73$\\
4    &    0.1 &   $5.\times 10^{-5}$ &   1   &  4    &   $0.0511$  		   & $2.92$\\
4    &    0.1 &   $5.\times 10^{-5}$ &   1   &  12  &   $0.153$  		   & $8.77$\\
10  &    0.1 &   $5.\times 10^{-5}$ &   1   &  1    &   $0.0319$  		   & $1.83$\\
10  &    0.1 &   $5.\times 10^{-5}$ &   1   &  12  &   $0.383$    		   & $21.9$\\
25  &    0.5 &   $5.\times 10^{-5}$ &   1   &  1    &   $0.179$   		   & $10.2$\\
4    &    0.5 &   $5.\times 10^{-5}$ &   1   &  1    &   $0.0286$   		   & $1.64$\\
4    &    0.1 &   $5.\times 10^{-5}$ &   0.1&  1    &   $0.00545$                  &$0.312$\\
4    &    0.5 &   $0.025$                   &   1  &   0.5 &   $2.86\times10^{-5}$ & $0.0323$ \\
25  &    0.5 &   $0.025$                   &   1  &   0.5 &   $1.78\times10^{-4}$ & $0.204$ \\ [1ex] 
\hline
\end{tabular}
\label{table}
\end{table}

We emphasise that the gradient/curvature drift velocity $V_g$ scales with the particle speed $v$ as $V_g/v \sim (r_g/L')(\delta B/B_0)$. In the unperturbed Parker spiral at a distance $1$ AU, if one uses a characteristic scale of spatial variation of $L'' \sim 1$ AU, by assuming that the gradient/curvature drift originates from the inhomogeneity of the Parker spiral itself, and not from the intermediate scale $L'$  ($L' \sim 0.001$ AU), $\Delta x _D$ would be suppressed by a factor $L'/L'' \simeq 0.001$. 
An extension to a turbulent Parker spiral magnetic field is underway and will be presented elsewhere.

\section{\label{Conclusion}Conclusion}

I have derived analytically the time-evolution of the average transverse displacement of a charged particle in a magnetostatic turbulence based on the critical balance GS95 conjecture. The particle motion is assumed to be well-approximated by the guiding-center motion. With no assumption of diffusion, I have calculated the perpendicular transport due to the single field line meandering, customarily regarded as dominant, and the gradient/curvature drift from the local perturbed field line. The former is dominated by the shear-Alfv\'en modes, the latter by the pseudo-Alfv\'en modes. Both contributions are found to be in free-streaming, not diffusive, regime. In particular, the departure of the guiding-center from the field line grows rapidly in time, invalidating the customary assumption that particles follow the field lines. If GS95 is the dominant MHD-scale turbulence in the solar wind, a crude estimate shows that  the longitudinal spread of energetic protons measured at 1 AU might be contributed to a non-negligible extent by cross-field transport due to gradient/curvature drift in the inhomogeneous turbulence, instead of the inhomogeneity of the unperturbed Parker spiral itself. The measured much larger (nearly $360^\circ$) longitudinal spread  certainly includes additional factors not considered here such as: angular extension and motion of the CME-driven shock producing the SEPs and Sun's spin dragging the Parker spiral across the heliosphere.

The perpendicular transport in turbulence presented here might imply for particle acceleration at quasi-perpendicular shocks an enhancement of the permanence time, hence of the maximal momentum, of energetic particles in the vicinity of the shock. Such a higher energy attainable combines with the small acceleration time at perpendicular shocks, as described in \citep{j87}. Turbulent field enhancements, generated downstream by inhomogeneities of the unshocked medium \citep{f13,f14}, can also increase the permanence time of particles at shocks. This result has relevant implications for particle acceleration at shocks, deferred to a separate work.

Finally, it is relevant to note that in this paper a particular model for the anisotropic power spectrum (Eq. \ref{Oughton_PS}) is assumed, that relies on the interpretation of solar wind measurements and  numerical simulations. However, the dominant process in the generation of anisotropy in MHD incompressible turbulence and the resulting turbulence power spectrum are currently topic of debate (see, e.g., \citet{gm10}). Moreover, a recent analysis of the magnetic field and plasma data from Wind spacecraft \citep{wthmw14} shows that the claimed spectral anisotropy in the solar wind inertial range can be explained by localised turbulent intermittency structures with no critical balance conjecture needed. Further experimental investigation is necessary to unveil the nature of the solar wind turbulence.

\acknowledgments

The author acknowledges useful discussions with M. Forman about the power spectrum modelling,  continuing discussions with J. R. Jokipii, J. K\'ota and J. Giacalone and useful suggestions from the two referees. The author benefitted from discussions at a workshop on ``First principles physics for charged particle transport in strong space and astrophysical magnetic turbulence'' held
at the International Space Science Institute (ISSI) in Bern, Switzerland. This work was supported by NASA under grants NNX13AG10G and NNX11AO64G.

\appendix

\section{\label{app_drift}Guiding-center drift}

In this section, I illustrate the details of the calculation of the non-vanishing terms for $d_D (t)$ in Sect. \ref{sect_drift}, which depends on the following components of $P_{ij} ({\bf k})$:
\begin{eqnarray}
P_{11} ({\bf k}) = \frac{k_2^2}{k_\perp^2} \Sigma({\bf k}) + \left( \frac{k_1 k_3}{k_\perp k} \right)^2 \Pi({\bf k})   ,\quad  P_{22} ({\bf k}) = \frac{k_1^2}{k_\perp^2} \Sigma({\bf k}) + \left( \frac{k_2 k_3}{k_\perp k} \right)^2 \Pi({\bf k}) \nonumber\\ 
P_{23} ({\bf k}) = -\frac{k_2 k_3}{k^2} \Pi({\bf k})  ,\quad P_{33} ({\bf k}) = \frac{k_\perp ^2}{k^2} \Pi({\bf k})  ,
\label{Pcomp}
\end{eqnarray}
where $k_\perp^2 = k_1^2 + k_2^2$ and $k_\parallel = k_3$. Analysis of Ulysses measurements \citep{wfho12} shows that only the tensor elements in Eq. \ref{Pcomp} contribute to the solar wind turbulence. We note that $P_{23}$ and $P_{33}$ contain only the power in the pseudo-Alfv\'en modes, whereas $P_{11}$ and $P_{22}$ are a combination of shear- and pseudo-Alfv\'en modes. 
  
{\bf Term 3322} - Using the last relation in Eq.\ref{Pcomp} and Eq.\ref{pseudo}, in cylindrical coordinates ($d^3 k = k_\perp d k_\perp d k_\parallel d \psi $) with $k_1 = k_\perp {\rm cos} \psi$, $k_2 = k_\perp {\rm sin} \psi$, $k_3 = k_\parallel$, it is readily found
\begin{eqnarray}
\int_{-\infty} ^{\infty} d^3 k  P_{33}({\bf k}) k_2^2 \frac{{\rm sin} [k_\parallel v_\parallel t]}{k_\parallel v_\parallel}  =  
   \frac{\varepsilon {\cal N} \pi B_0^2}{v_\parallel \ell^{1/3}} \int_{k_\parallel ^0} ^{k_\parallel ^{M}} d k_\parallel  \frac{{\rm sin} [k_\parallel v_\parallel t]}{k_\parallel }  I_{5/3}(k_\parallel, k_\perp ^0, k_\perp ^{M})  
\label{drift_gen3322}
\end{eqnarray}
where $k_\perp ^0, k_\perp ^{M}$ are the outer and the dissipation perpendicular wave-numbers respectively; the boundaries $k_\parallel ^0, k_\parallel ^{M}$ are related to $k_\perp ^0, k_\perp ^{M}$ through critical balance scaling: $k_\parallel ^0 \simeq (k_\perp^0)^{2/3} / L^{1/3}$, $k_\parallel ^M \simeq (k_\perp^M)^{2/3} / L^{1/3}$ for $M_A =1$. We consider here (FJ11) perpendicular transport on time-scale shorter than $1/k_\parallel v_\parallel t $. In Eq. \ref{drift_gen3322} the auxiliary function
\begin{equation}
I_{\alpha}(k_\parallel, k_\perp ^0, k_\perp ^{M})  = \int_{k_\perp ^0} ^{k_\perp ^{M}} d k_\perp  \frac{k_\perp^{\alpha}}{k_\perp^{2} + k_\parallel ^2 } {\rm exp}\left( \frac{-\ell^{1/3} k_\parallel}{k_\perp^{2/3}}\right)
\label{idef}
\end{equation}
was defined. The expansion in series of the exponential factor in $I_{5/3}(k_\parallel, k_\perp ^0, k_\perp ^{M}) $ by using ${\rm e}^{-x} = \sum_{n=0}^{\infty} (-x)^n/{n!}$ yields 
\begin{eqnarray}
I_{5/3}(k_\parallel, k_\perp ^0, k_\perp ^{M})  \simeq \sum_{n=0} ^{3}  \frac{(-\ell)^{n/3}}{n!}  \frac{3 k_\parallel^{n-2}k_\perp^{2\frac{4-n}{3}}}{2(4-n)} \times 
{}_{2}F_{1} \left.\left(1,\frac{4-n}{3};\frac{7-n}{3};-\frac{k_\perp^2}{k_\parallel^2}\right)  \right|_{k_\perp ^0}^{k_\perp ^{M}} 
\label{iseries}
\end{eqnarray}
where ${}_{2}F_{1} (a,b;c;z)$ is the ordinary hypergeometric function. In Eq. \ref{iseries} the sum is truncated at $n=3$ (\citet{gr73}, Eq. 3.194.5, applies in Eq. \ref{iseries} only for $n<4$); in what follows it is shown that it suffices to consider the term $n = 0$.

The right hand side of Eq. \ref{drift_gen3322} can then be approximated by using the expansion in Eq.\ref{iseries} as
\begin{flalign}
&  \frac{\varepsilon {\cal N}\pi B_0^2}{v_\parallel \ell^{1/3} L^{2/3}}   \sum_{n=0} ^{3}  \frac{3 (-\ell/L)^{n/3}}{n! 2(4-n)} \left (\frac{v_\parallel t}{L} \right)^{2-n} \times & \nonumber \\
&  \left[ (k_\perp L)^{2\frac{4-n}{3}} \times   
{}_{2}F_{1} \left.\left(1,\frac{4-n}{3};\frac{7-n}{3};-(k_\perp L)^{2/3} \right) \right] \right|_{k_\perp ^0}^{k_\perp ^{M}} S_n(x^0, x^M) &
\label{series_tot_pre}
\end{flalign}
where $x = k_\parallel v_\parallel t$ and $S_n(x_0, x_M) = \int_{x_0}^{x_M} dx ~x^{n-3} {\rm sin} x$. In Eq.\ref{series_tot_pre} the critical balance condition ($k_\parallel \sim k_\perp^{2/3}/ L^{1/3}$) is used; thus, the independent variable of ${}_{2}F_{1} (\cdot)$ in Eq.\ref{iseries} can be approximated as $-(k_\perp/k_\parallel)^2 \sim (k_\perp L)^{2/3}$ and likewise for the boundaries of $x$-integration, i.e., $x_0 = k_\parallel ^0 v_\parallel t \sim (k_\perp^0)^{2/3} v_\parallel t /L^{1/3}$ and $x_M \sim (k_\perp^M)^{2/3} v_\parallel t /L^{1/3}$.

The argument for the truncation at $n=3$ in Eq. \ref{iseries} goes as follows. The integrals $S_n(x_0, x_M)$ are extended over a time shorter than $1/k_\parallel v_\parallel t $, i.e., $x_M < 1$, and in the interval $(x_0, x_M)$ I have $S_0 \simeq \left[{{\rm Si}(x)\over 2} + {{\rm sin}(x)\over 2x^2} + {{\rm cos}(x)\over 2x}\right] |_{x_0}$, $S_1 \simeq \left[{{\rm sin}(x)\over x}  - {\rm Ci}(x) \right] |_{x_0}$ , $S_2 \simeq {\rm Si}(x_M) \simeq \pi/2$ and $S_3 = {\rm cos} (x_0) \simeq 1$, where ${\rm Si} (x)$ and ${\rm Ci}(x)$ are respectively Sine and Cosine integrals. It is easily shown that it holds $|S_n(x^0, x^M)/S_0(x^0, x^M)| \ll 1$ (for $n>1$). Moreover, it is possible to show for the coefficients $A_n (k_\perp L)= (k_\perp L)^{2\frac{4-n}{3}} \times {}_{2}F_{1} \left(1,\frac{4-n}{3};\frac{7-n}{3};-(k_\perp L)^{2/3} \right)$ in Eq.\ref{series_tot_pre} that $A_1/A_0$, $A_2/A_0$ and $A_3/A_0$ are rapidly decreasing functions of $k_\perp L $ over at least three decades in $k_\perp L$. Finally, the $n-$th term in Eq. \ref{series_tot_pre} evolves in time as $t^{2-n}$. This justifies retaining only the term $n=0$ in Eq. \ref{series_tot_pre}:
\begin{align}
\frac{3}{8} \frac{\varepsilon {\cal N}\pi B_0^2}{v_\parallel \ell^{1/3} L^{2/3}}  {(v_\parallel t )^2 \over {L^{2}}}  \left.\left[  {{\rm Si}(x)\over 2} + {{\rm sin}(x)\over 2x^2} + {{\rm cos}(x)\over 2x}\right] \right|_{x^0} 
\left. \left[(k_\perp L)^{8/3}  {}_{2}F_{1} \left(1,{4\over 3}; {7 \over 3};- (k_\perp L)^{2/3} \right) \right] \right|_{k_\perp ^{M}} \nonumber.
\end{align}
By using the fact that $\left[{{\rm Si}(x)\over 2} + {{\rm sin}(x)\over 2x^2} + {{\rm cos}(x)\over 2x}\right] |_{x^0} \sim 1/x_0$ for $x_0 \ll 1$, the right hand side of Eq. \ref{drift_gen3322} can then be approximated as
\begin{align}
 \frac{3 \varepsilon {\cal N}\pi B_0^2 t}{8 k_\parallel ^0 \ell^{1/3} L^{8/3}} 
\left. \left[(k_\perp L)^{8/3}  {}_{2}F_{1} \left(1,{4\over 3}; {7 \over 3};- (k_\perp L)^{2/3} \right) \right] \right|_{k_\perp ^{M}} 
\label{series_tot_1}
\end{align}
The previous calculation shows that the term in Eq. \ref{drift_gen3322} grows linearly in time. The corresponding pitch-angle factor $F(\mu^2) = ((1-\mu^2)/2)^2$, averaged over an isotropic pitch-angle distribution, gives a factor $2/15$.

{\bf Term 2323} - Using the expression for $P_{23}$ in Eq.\ref{Pcomp} and Eq.\ref{pseudo}, in cylindrical coordinates, it is found  
\begin{eqnarray}
\int_{-\infty} ^{\infty} d^3 k  P_{23}({\bf k}) k_2 k_\parallel \frac{{\rm sin} [k_\parallel v_\parallel t]}{k_\parallel v_\parallel}  =  
 -  \frac{\varepsilon {\cal N}\pi B_0^2}{v_\parallel \ell^{1/3}} \int_{k_\parallel ^0} ^{k_\parallel ^{M}} d k_\parallel {k_\parallel } {{\rm sin} [k_\parallel v_\parallel t]}  I_{-1/3}(k_\parallel, k_\perp ^0, k_\perp ^{M})  ,
\label{drift_gen3223}
\end{eqnarray}
where $I_\alpha$ is defined in Eq. \ref{idef}. 
Expanding in series $I_{-1/3} (k_\parallel, k_\perp ^0, k_\perp ^{M}) $ I find that the right hand side of Eq.\ref{drift_gen3223} can be cast as
\begin{equation}
-  \frac{\varepsilon {\cal N}\pi B_0^2}{v_\parallel \ell^{1/3}}   \sum_{n=0}^{\infty}  \frac{ (-L)^{n/3}}{n!}  \frac{1}{(v_\parallel t)^n} \left[  \int_{x^0}^{x^M} dx ~x^{n-1} {\rm sin} x \right]  \int_{k_\perp^0}^{k_\perp^M} dk_\perp \frac{k_\perp ^{-\frac{2n+1}{3}}}{k_\perp^2 + k_\parallel^2} .
\label{i'series}
\end{equation}
In Eq. \ref{i'series} the $n^{th}$ term decreases as $(v_\parallel t)^{-n}$; moreover, the integrals $\int_{x_0}^{x_M} dx ~x^{n-1} {\rm sin} x$ for $x_M <1$ and the integrals in $k_\perp$ are decreasing functions of $n$. Therefore the series in Eq. \ref{i'series} can be approximated to the lowest order, $n=0$ (\citet{gr73}, Eq. 3.194.5, valid for $n<1$). Using again the critical balance condition ($k_\parallel \sim k_\perp^{2/3}/ L^{1/3}$) leads to approximate the right hand side of Eq.\ref{drift_gen3223} as:  
\begin{equation}
- \frac{3 \varepsilon {\cal N}\pi B_0^2}{2 v_\parallel \ell^{1/3}L^{2/3}} [  {{\rm Si}(x)} ] |_{x^0}  {3\over 2} (k_\perp^M L)^{2/3}  {}_{2}F_{1} \left.\left(1,{1\over 3}; {4 \over 3};- (k_\perp L)^{2/3} \right) \right|_{k_\perp ^{M}} 
\label{series_tot_2}
\end{equation}
The only time-dependence of Eq. \ref{series_tot_2} is the slow variation in the factor ${{\rm Si}(x)} ] |_{x^0} $. The ${}_{2}F_{1}$ coefficient in Eq. \ref{series_tot_2} is much smaller than in Eq. \ref{series_tot_1}. Thus, the term $(2323)$ in the drift transport coefficient is neglected.

{\bf Term 2233} - The last non-vanishing term for the instantaneous drift transport coefficient includes contribution from both shear- and pseudo-Alfv\'en modes. From the expression of $P_{22}$ in Eq.\ref{Pcomp} I find 
\begin{eqnarray}
\int_{-\infty} ^{\infty} d^3 k  P_{22}({\bf k}) k_\parallel^2 \frac{{\rm sin} [k_\parallel v_\parallel t]}{k_\parallel v_\parallel}  =  
 \frac{{\cal N}\pi B_0^2}{v_\parallel } \int_{k_\parallel ^0} ^{k_\parallel ^{M}} d k_\parallel {k_\parallel } {{\rm sin} [k_\parallel v_\parallel t]} \times \nonumber\\ \left( \underbrace{{1\over L^{1/3}}J(k_\parallel, k_\perp ^0, k_\perp ^{M})}_{\rm shear} - \underbrace{{\varepsilon \over \ell^{1/3}} I_{-7/3}(k_\parallel, k_\perp ^0, k_\perp ^{M})}_{\rm pseudo} \right)  
\label{drift_gen2233}
\end{eqnarray}
where the $J$-term accounts for shear-modes and $I_{-7/3}$ for the pseudo-. The auxiliary function $J$ is given by
\begin{eqnarray}
J(k_\parallel, k_\perp ^0, k_\perp ^{M}) & = & \int_{k_\perp ^0} ^{k_\perp ^{M}} d k_\perp   k_\perp^{-7/3} {\rm exp}\left( \frac{-L^{1/3} k_\parallel}{k_\perp^{2/3}}\right) \nonumber \\
&\simeq & \frac{3}{2k_\parallel L^{1/3}} \left( \frac{1}{k_\parallel L^{1/3}} + \frac{1}{{k_\perp^M }^{2/3}}\right) {\rm exp}\left( \frac{-L^{1/3} k_\parallel}{{k_\perp^M}^{2/3}}\right) 
\label{jdef}
\end{eqnarray}
where in the second equivalence the contribution from $k_\perp^0$ is neglected. Expanding again in series the exponential in the last member of Eq. \ref{jdef}, I find that the shear term in Eq. \ref{drift_gen2233} can be cast as
\begin{equation}
\frac{3 {\cal N}\pi B_0^2}{ v_\parallel L^{2/3}}   \sum_{n=0}^{\infty}  \left(\frac{ -L^{1/3}}{{k_\perp^M}^{2/3}}\right)^n \frac{1}{n!} \frac{1}{(v_\parallel t)^{n+1}} \left[  \int_{x^0}^{x^M} dx ~x^n {\rm sin} x \right]  \left( \frac{v_\parallel t}{x L^{1/3}} + \frac{1}{{k_\perp^M }^{2/3}}\right) 
\label{series_tot_3}
\end{equation}
where the $n^{th}$ term decreases as $t^{-n}$. Thus, to the lowest order, $n=0$, Eq.\ref{series_tot_3} is approximated by $3{\cal N}\pi^2 B_0^2/(2 v_\parallel L) $ (by using ${\rm Si}(x_M) \simeq \pi/2$), independent on time, so negligible with respect to the term $(3322)$ for the reason explained in this Appendix.
As for the pseudo modes, the $I_{-7/3}$-term in Eq.\ref{drift_gen2233} can be approximated, using again the series expansion of the exponential, by
\begin{equation}
\frac{\varepsilon {\cal N} \pi B_0^2}{v_\parallel \ell^{1/3}}\sum_{n=0} ^{\infty} \frac{(-\ell)^{n/3}}{n!} (v_\parallel t)^{-\frac{n+2}{3}} \int_{x^0} ^{x^{M}} dx ~{\rm sin}x ~x^{\frac{n-1}{3}}  \int_{y_0}^{y_M} dy \frac{y^{-\frac{7+2n}{3}}}{1+y^2}
\label{series_tot_4}
\end{equation}
where $y=k_\perp/k_\parallel \simeq (k_\perp L)^{1/3}$. The integrals in $y$ for $n=0,1,2$ over three decades ($y_0>1$) are of order ${\cal O}(1)$.  The dominant term ($n=0$) falls in time as $t^{-2/3}$. We conclude that this term is negligible compared to the term $(3322)$.

\section{\label{app_MFL}Magnetic field line random walk}

In this section, I illustrate the details of the calculation of $d_{MFL} (t)$. By using the definition of $P_{11}$ in the first relation of Eq.\ref{Pcomp} (the result is unchanged if $P_{22}$ given by the second relation in Eq.\ref{Pcomp} is used) and Eq.\ref{dMFL2} it is found in cylindrical coordinates ($d^3 k = k_\perp d k_\perp d k_\parallel d \psi $) with $k_1 = k_\perp {\rm cos} \psi$, $k_2 = k_\perp {\rm sin} \psi$, $k_3 = k_\parallel$
\begin{eqnarray}
  \frac{1}{B_0^2} \int_{-\infty} ^{\infty} d^3 k  P_{11}({\bf k}) \frac{{\rm sin} [k_\parallel v_\parallel t]}{k_\parallel}  = 
{\cal N}\pi  \int_{k_\parallel ^0} ^{k_\parallel ^{M}} d k_\parallel {{\rm sin} [k_\parallel v_\parallel t]}  \times\nonumber\\ 
\left( \underbrace{{1\over k_\parallel L^{1/3}}J(k_\parallel, k_\perp ^0, k_\perp ^{M})}_{\rm shear} + \underbrace{{\varepsilon k_\parallel \over \ell^{1/3}} I_{-7/3}(k_\parallel, k_\perp ^0, k_\perp ^{M})}_{\rm pseudo} \right)  
\label{MFL_tot}
\end{eqnarray}
where $J$ and $I_{\alpha}$ are defined respectively in Eq.s \ref{jdef} and \ref{idef}. The $J$-term, from the second relation in Eq. \ref{jdef}, is given by
\begin{equation}
\frac{3\pi {\cal N}}{2 L^{2/3}}   \int_{k_\parallel ^0} ^{k_\parallel ^{M}} d k_\parallel \frac{{\rm sin} [k_\parallel v_\parallel t]}{k_\parallel ^2}  \left( \frac{1}{k_\parallel L^{1/3}} + \frac{1}{{k_\perp^M }^{2/3}}\right) {\rm exp}\left( \frac{-L^{1/3} k_\parallel}{{k_\perp^M}^{2/3}}\right) .
\end{equation}
By using the series expansion of the exponential and the same approximation as in Eq. \ref{series_tot_3}, the shear-modes contribution to MFL transport coefficient can be approximated to the lowest order by 
\begin{equation}
({\rm shear}) \simeq \frac{3\pi {\cal N}}{2 } \frac{v_\parallel t}{k_\parallel^0  L} 
\label{shear_MFL_0}
\end{equation}
where the approximation $\left[{{\rm Si}(x)\over 2} + {{\rm sin}(x)\over 2x^2} + {{\rm cos}(x)\over 2x}\right] |_{x^0} \sim 1/x_0$ for $x_0 \ll 1$ was used. Thus, the shear term grows linearly in time. The pseudo-modes term in Eq. \ref{MFL_tot} is equal to the pseudo-modes term in Eq. \ref{drift_gen2233}, modulo a constant factor $-B_0^2/v_\parallel$; due to its decay in time for every $n$, it is neglected here with respect to the shear-modes term. 

\end{document}